# Mysterious non-detection of HeI (2$^3$S) transit absorption of GJ436b

M. S. Rumenskikh[1,2,3], M. L. Khodachenko[4], I. F. Shaikhislamov[1,3], I. B. Miroshnichenko[1,5],

A. G. Berezutsky[1], A. V. Shepelin[1], N. K. Dwivedi[6]

[1] *Institute of Laser Physics SB RAS, Novosibirsk, Russia*
[2] *Pushkov Institute of Terrestrial Magnetism, Ionosphere and Radiowave Propagation of the Russian Academy of Sciences (IZMIRAN), Troitsk, Moscow 108840, Russia*
[3] *Institute of Astronomy RAS, Moscow, Russia*
[4] *Space Research Institute, Austrian Academy of Sciences, 8040 Graz, Austria*
[5] *Novosibirsk State Technical University, Novosibirsk, Russia*
[6] *Institute of Earth Physics and Space Science, 9400 Sopron, Hungary*

**Abstract:** Possible reasons for the non-detection of absorption in the metastable HeI(2$^3$S) line at transit observations of warm Neptune GJ436b, in spite of the well pronounced strong absorption features measured earlier in Lyα for this planet, are investigated. We perform numeric simulations of the escaping upper atmosphere of this planet and its HeI(2$^3$S) triplet absorption with a global 3D multi-fluid self-consistent hydrodynamic model. By fitting the model parameters to the lowest detection level of absorption measurements, we constrain an upper limit the He/H abundance three times smaller than the solar value. We demonstrate that neither the significant changes of the stellar wind related with possible stellar coronal mass ejections (CMEs), or possible variations in the stellar ionization radiation, nor the presence of heavy trace elements have crucial effect on the absorption at the 10830Å line of HeI(2$^3$S) triplet. The main reason of weak signature is that the region populated by the absorbing metastable helium is rather small (<3Rp), as well as the small size of the planet itself, in comparison to the host star. We show that the radiation pressure force acting on the HeI(2$^3$S) atoms spreads them along the line of sight and around the planet, thus further reducing peak absorption.

**Introduction**

The era of Hubble Space Telescope observations gave numerous insights into the physics of exoplanetary atmospheres revealing the existence of extended hydrogen envelopes around a number of hot Jupiters and warm Neptunes. Measurements in Lyα line yielded the transit depths of ~15% (*Vidal-Madjar et al. 2003*), ~50% (*Ehrenreich et al. 2015*), and 35% (*Bourrier et al. 2018*), for HD209458b, GJ436b, and GJ3470b correspondingly. Spectrally-resolved observations of GJ436b and GJ3470b have shown significant absorption at velocities ~100 km/s in blue wing of Lyα line, indicating on the generation of Energetic Neutral Atoms (ENAs) from Stellar Wind (SW) protons in interaction with much slower atmospheric particles (*Holmstrong et al. 2008, Khodachenko et al. 2019*). Moreover, the GJ436b case is specially prominent because the measurements performed with high temporal and spectral resolution have shown significantly earlier ingress and much longer egress in Lyα line than those observed in white light (*Lavie et al. 2017*). This indicates on the existence of an extended hydrogen tail which covers a significant part of the orbital trajectory of the planet. As proposed in *Lammer et al. (2003),* and studied thereafter in many works, the upper atmosphere expansion and escape are caused by radiative heating and ionization of the planetary atmospheric material by stellar soft X-ray and EUV (so-called XUV) radiation. The evidence of expanding hydrogen atmospheres of the mentioned hot exoplanets makes it reasonable to expect that similarly well pronounced absorption in helium lines might take place and be detected. However, among about 30 planets observed in HeI($2^3$S) line, the absorption was detected only for about half, with several cases obviously baffling the expectations (*Fossati et al. 2023*).

The potential importance of the metastable helium absorption measurements for probing of exoplanetary atmospheres was first pointed in *Seager & Sasselov (2000)*, who predicted a possibility of HD209458b transit observations in HeI($2^3$S) line ($\lambda$=10830Å) in relation to the escaping atmospheric material. However, the estimates for specific planetary systems were implemented later. *Oklopčić & Hirata (2018)* using 1D hydrodynamic Parker-wind two-component model of exoplanetary atmospheres, were first to calculate and predict the HeI($2^3$S) absorption for HD209458b and GJ436b at the level of 2% and 8%, respectively. Later, this forecast was partially confirmed with observations. In particular, *Alonso-Floriano et al. 2019* reported for HD209458b the detection of transit absorption at the position of HeI($2^3$S) triplet at the level of 0.9 ± 0.10%. However, similar observations of GJ436b revealed a surprising non-detection of the absorption in HeI($2^3$S) line (*Nortmann et al. 2018*). This unexpected observational fact remains explained and has to be physically justified.

There might be several possible reasons for the lack of transit absorption in HeI($2^3$S) line at GJ436b. First, the planet may contain no helium at all in the upper atmosphere, depending on weather it is of primordial or secondary nature. The primordial atmospheres retain helium from the accretion disk, along with the heavy trace elements (e.g., carbon, magnesium, oxygen, silicon and etc.) in the related to hydrogen abundance of He/H~0.1. In contrast, the secondary atmospheres consist of material outgassed from the interior, so they are believed to be dominated by heavy elements with the chemically bound hydrogen, while helium appears to be mostly lost at earlier stages. Therefore, the hypothesis of an absent helium can be circumstantially verified with the measurements of transit absorption in the lines of heavy elements and/or their molecular compounds.

Interestingly, the question of heavy elements in the upper atmosphere of GJ436b poses a similar problem as helium. Based on Spitzer photometry *Lanotte et al. (2014)* concluded that the lower atmosphere which start with pressures ~$10^{-4}$ bar (see figure 1 from *Madhusudhan 2019*) is metal reach and enhanced in CO and $CO_2$. However, the reanalysis of available HST-COS FUV observations by *Santos et al. 2019* have not found absorption in the strongest FUV stellar lines of C II, C III, Si II, Si III, Si IV, NV, OV. The non-detection of absorption in CII and SiIII lines is also reported by *Loyd et al. 2017*. 1D aeronomy simulation of *Loyd et al. 2017* yielded 2% for the line integrated transit depth in CII, which is well below the data-imposed upper limits. The seeming absence of metals in the upper atmosphere of GJ436b led Santos et al. 2019 to suggest that the mixing is not efficient in dragging the Si-rich clouds high enough to be caught in the hydrodynamic escape.

Additionally, the varying stellar activity might be of importance, since the upper atmosphere and transmission spectra are affected by fluctuations of the stellar radiation. Starspots, stellar coronal mass ejections (hereafter CMEs), or varying stellar emission, could contaminate the transit absorption profiles (*Santos et al. 2019*). However, as argued in *Llama et al. 2016* and *Cauley et al. 2018,* these are not the case for the transits of GJ436b observed in Ly$\alpha$ and HeI($2^3$S) lines.

Finally, the above mentioned model by *Oklopčić & Hirata 2018* used for the prediction of HeI($2^3$S) absorption, has a number of crucial simplifications. It is one-dimensional, and includes only basic photochemical reactions involving helium and hydrogen. This model cannot reproduce the complex 3D dynamics of upper atmosphere which plays a key role in formation of the transmission spectroscopy features (e.g. spectral absorption profiles and light curves). The 3D hydrodynamic (HD) model, we use in this study to interpret absorption of the stellar NIR flux in the HeI($2^3$S) line during the transits of GJ436b, has been already applied for the simulation and interpretation of observations of spectrally-resolved transits in Ly$\alpha$ of this planet (*Khodachenko et al. 2019*). The SW density, temperature and speed at the planetary orbit, stellar XUV radiation intensity and other parameters obtained in *Khodachenko et al. 2019* by fitting of the simulated absorption features to observations, are not only physically justified and reasonable, but appear as well in agreement with other simulation results (*Villarreal D'Angelo et al. 2021*). Our 3D HD model was also successfully applied for the interpretation of spectroscopic observations of other exoplanets, such as $\pi$ Men c (*Shaikhislamov et al.2020a*), HD209458b (*Shaikhislamov et al. 2020b; Khodachenkoet al. 2021b*), GJ3470b (*Shaikhislamov et al. 2020c*), Wasp107b (*Khodachenko et al. 2021a*), and HD189433b (*Rumenskikh et al. 2022*). These studies included also the interpretation of absorption at the position of metastable helium triplet.

The paper is structured as follows. In section 2 we provide a short description of our model and details regarding the simulation parameters used. Section 3 describes our findings concerning the interpretation of absorption non-detection in the HeI($2^3$S) line. In particular, we analyze there, how the stellar CMEs, XUV intensity, the radiation pressure, atmospheric composition could affect the transit absorption. The discussion of the modelling results is presented in Section 4.

## 2. The model and parameters of simulation

Our 3D HD model, introduced in *Shaikhislamov et al. (2018a)* was developed from the previous 1D (*Shaikhislamov et al. 2014*) and 2D (*Shaikhislamov et al. 2016, 2018b, Khodachenko et al. 2015, 2017*) models. In the present paper we simulate the HeI($2^3$S) transit absorption of this planet and analyze its detectability. All the basic modelling approaches and equations for calculating the metastable helium level population and depopulation were addressed in our earlier work devoted to the HeI($2^3$S) triplet absorption at GJ3470b (*Shaikhislamov et al. 2020*). Therefore, in this section we provide only the most important details regarding our model and specify the used parameters. Further relevant details can be found in *Shaikhislamov et al. (2020), Khodachenko et al. (2019),* and *Rumenskikh et al. (2022)*.

The HD model solves the continuity, momentum and energy equations using an up-wind explicit second-order accuracy scheme in a global non-inertial spherical frame of reference, attached to the planet and orbiting with it around the host star. The center of coordinates in this frame of reference coincides with the planet center, and X-axis is always pointed to the star, so it could be called as a "tidally locked" frame of reference. The escaping planetary wind (PW) is driven by the heating and ionizing impact of the stellar XUV radiation absorbed by the planetary upper atmospheric gas. Altogether, the model accounts for such important factors as the radiation pressure, acting on hydrogen and helium atoms, cooling and heating by heavy elements (if present), generation of ENAs due to charge exchange between stellar protons and planetary hydrogen atoms, as well as the set of hydrogen-helium photochemistry reactions (*Khodachenko et al. 2015; Shaikhislamov et al. 2016, 2021*).

Besides of hydrogen and helium, our model also includes the dynamics of such heavy elements as oxygen, carbon, and silicon. It is assumed that the abundances of heavy elements are close to the solar-like ones. Therefore, due to the relatively low content, these elements remain as passive admixtures, not affecting the global dynamics of the planetary upper atmosphere.

The free model parameters include the pressure, 0.05 bar, and the temperature, T=750 K, at the base of upper atmosphere (the latter is taken close to that measured for GJ436b by *Deming et al. (2007)* during the secondary transit, 717 K). The measured near-IR radiation flux around the 10830 Å line of the metastable helium triplet is taken as $F_{1083}$=3 erg·cm$^{-2}$·s$^{-1}$·Å$^{-1}$, which determines the radiation pressure, acting on the metastable helium atoms. To study the influence of space weather factors on HeI($2^3$S) absorption, the ionizing stellar radiation flux $F_{XUV}$ ($\lambda$<912 Å), and the necessary for SW modelling (for details see in *Khodachenko et al. 2019*) stellar mass loss rate, $M'_{sw}$, were varied in different model runs, whereas other parameters of the SW model, such as coronal temperature, $T_{cor}$ = 3·10$^6$ K and the terminal speed, $V_{SW,\infty}$= 540 km/s were kept unchanged. All these parameters are constrained by the performed earlier work of *Khodachenko et al. (2019)* and *Villarreal D'Angelo (2021)* model-based interpretation of the in-transit Lyα absorption of GJ436b. Altogether, depending on the used value of the stellar mass loss rate, we distinguish between different types of the SW, specifically, "weak" $M'_{sw}$ =2.5·10$^{11}$g/s, which is ten times smaller than the corresponding solar value, "medium", $M'_{sw}$ =2.5·10$^{12}$g/s, and "strong", $M'_{sw}$ =2·10$^{13}$g/s. It is worth mentioning that according to *Khodachenko et al. (2019)* the best correspondence between the simulated and measured Lyα absorption was achieved with the "weak" SW. Besides of that, to evaluate the influence of other physical factors on the HeI($2^3$S) absorption, we varied in dedicated model runs the value of radiation pressure, by assuming different stellar near-IR fluxes, and the initial atmospheric composition, by assuming atomic, instead of molecular, hydrogen atmosphere or adding heavy trace elements. The parameters and outcomes of all these different simulations aimed to explain the non-detection of the excess transit absorption in HeI($2^3$S) line, are summarized in Table 1. The details of calculating absorption in lines of CII and SiIII are presented in our earlier papers (*Shaikhislamov et al. 2018b, 2020*), while that of HeI($2^3$S) line in (*Shaikhislamov et al. 2021b, Khodachenko et al. 2021b, Rumenskikh et al. 2022*).

| No. | $F_{XUV}$, erg·cm$^{-2}$·s$^{-1}$ | $F_{10830}$, erg·cm$^{-2}$·s$^{-1}$·Å$^{-1}$ | $M'_{sw}$ | He/H | $M'_{pw}$ | FWHM HeI($2^3$S), km/s | Abs HeI($2^3$S), % | Abs CII, % | Abs SiIII, % | other |
|---|---|---|---|---|---|---|---|---|---|---|
| 1 | 0.86 | 3 | 25 | 0.03 | 0.59 | 12 | 0.74 | | | |
| 2 | 0.86 | 3 | 100 | 0.03 | 0.57 | 12.2 | 0.83 | | | |
| 3 | 0.86 | 3 | 200 | 0.03 | 0.56 | 11.9 | 0.75 | | | |
| 4 | 0.86 | 3 | 400 | 0.03 | 0.57 | 10.2 | 0.64 | | | |
| 5 | 0.86 | 3 | 2000 | 0.03 | 0.63 | 9.7 | 0.42 | | | |
| 6 | 0.43 | 3 | 25 | 0.03 | 0.26 | 10.4 | 0.57 | | | |
| 7 | 1.7 | 3 | 25 | 0.03 | 1.29 | 12.5 | 1.06 | | | |
| 8 | 0.86 | 0.3 | 25 | 0.03 | 0.58 | 11.6 | 1.2 | | | |
| 9 | 0.86 | 20 | 25 | 0.03 | 0.6 | 12.1 | 0.36 | | | |
| 10 | 0.86 | 3 | 25 | 0.03 | 3.02 | 14.6 | 3.4 | | | Atomic atm. |
| 11 | 0.86 | 3 | 25 | 0.03 | 0.52 | 13.4 | 0.4 | 1.5 | 2.5 | tr.el., x0.3 Sol |
| 12 | 0.86 | 3 | 25 | 0.1 | 0.75 | 12.7 | 1.3 | 1.1 | 1.7 | tr.el., x1 Sol |
| 13 | 0.86 | 3 | 25 | 0.03 | 0.6 | 11.3 | 0.39 | 3.0 | 3.5 | tr.el., x2 Sol |
| 14 | 0.86 | 3 | 25 | 0.2 | 0.8 | 15.5 | 2.2 | | | tr.el., x2 Sol |
| 15 | 0.86 | 3 | 25 | 0.015 | 0.6 | 12.0 | 0.36 | | | |
| Observations | | | | | | | ~0.41, [1] | 4.8, [2] | 7.8, [2] | |

**Table 1.** The parameter sets of discussed model runs. Columns from left to right: number of the model run; stellar XUV flux in the range of wavelengths 10<λ<912 Å scaled to a reference distance of 1 a.u; stellar flux near the wavelength λ<10830 Å at 1 a.u; the stellar mass loss $M'_{sw}$ rate expressed in 10$^{10}$ g/s; helium abundance in the base atmosphere; calculated mass loss rate $M'_{pw}$ of the planet atmosphere; half-maximum width and maximum of the simulated HeI($2^3$S) excess absorption profile at mid-transit; integrated over lines absorption in CII and SiIII, calculated

for the corresponding simulated profiles at mid-transit. The additional conditions applied in some model runs are indicated in the last column, e.g., atomic hydrogen atmospheric content; abundance of heavy trace elements (C, Si) being either Solar (x1 Sol), reduced (x0.3 Sol), and increased (x2 Sol). The last row contains the observationally measured upper limits of non-detection according to [1] - *Nortmann et al. (2019)* and [2] - *Dos Santos et al. (2017)*.

## 3. HeI($2^3$S) absorption modelling under different space weather conditions

### *3.1 The influence of SW parameters and XUV flux on the structure of absorbing area*

Previous interpretations of the Lyα transit absorption of GJ436b allow constraining such important parameters of space weather around the planet as stellar XUV radiation flux, as well as velocity, density and temperature of the SW. These parameters were derived assuming the standard solar-like abundance of helium, while in this work we consider it as a value to be constrained by observations. At the same time, according to the study reported in *Khodachenko et al. 2019* (e.g., Fig.14 there), the influence of helium content on the shape of Lyα absorption profiles during the transit is relatively weak. In particular, the variation of He/H in the range between 0.1 and 1 considered in *Khodachenko et al. (2019)*, results in the variation of the Lyα transit absorption depth within the measurement error. Therefore, in our study of the GJ436b transit absorption at the position of HeI($2^3$S) triplet, we take as a basic set of model parameters those space weather conditions ($F_{XUV}$=0.86 erg·cm$^{-2}$·s$^{-1}$ at 1 a.u., and $M'_{sw}$= 25 x 10$^{10}$ g/s), which were justified in *Khodachenko et al. 2019* by fitting of the Lyα transit measurement, but for the abundance of helium He/H we consider a broader range, from 0.015 to 0.2. This range of helium abundances includes also the values much smaller than those in the Solar system, predicted by models based on observations for a number of planets (e.g., in *Lampón et al. 2020, 2021*).

To understand the interplay of different photochemistry processes responsible for the population and depopulation of the metastable HeI($2^3$S) level in the expanding upper atmosphere of GJ436b, we compare in Figure 1 the corresponding reaction rates for the basic case of the model run N1. The radiative transition between ground and metastable level is strongly suppressed, and recombination of HeII with free electrons (reaction №9) becomes the main source of the $2^3$S level pumping (*Shaikhislamov et al. 2020c, Khodachenko et al. 2021a*). It is balanced mostly by the auto-ionization collisions with H and H$_2$ particles (reactions № 7, 8) and the electron impact transfer from the triplet to singlet state (reactions №3,4). Photoionization rate from the $2^3$S state (reaction №2) is relatively small. At the altitudes >3$R_p$, the metastable HeI($2^3$S) population becomes too low to produce absorption due to exponential decrease of HeII density. In the shock region (~11$R_p$) collisional excitation by hot electrons (reaction № 6) becomes the main pumping process, however it does not contribute much to the absorption due to small density of HeI in this region.

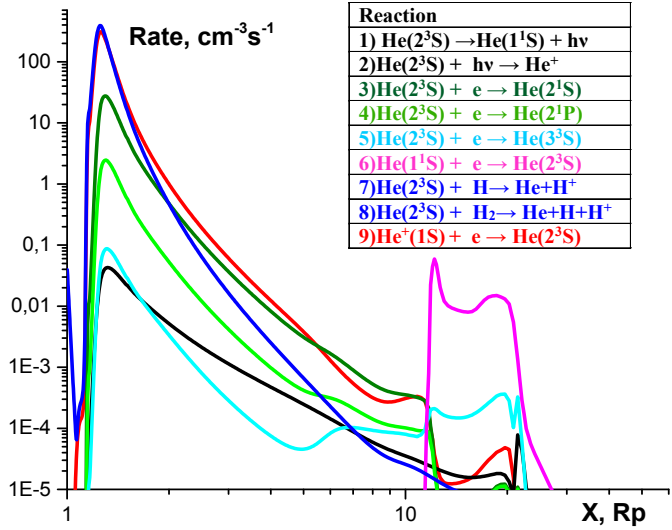

**Figure 1.** Rates of the reactions responsible for the population and depopulation of the metastable HeI($2^3$S) level versus the distance along the planet-star line (axis X) under conditions of the model run N1. The colored plot lines represent sums of reactions, marked in the plot legend with the same color.

As any star, the host of GJ436b might generate coronal mass ejections (CMEs), which can influence the transmission spectroscopy measurements. To check this possibility and to reveal how the SW conditions affect the absorbing HeI($2^3$S) fraction, we performed a series of model runs with different stellar mass loss rates $\dot{M}_{sw}$ (N1-N5, Table 1), which are proportional to the SW density, while keeping all other parameters of the model the same. Left panel in Figure 2 shows the distributions of the main physical parameters of the escaping PW along the planet-star line (X-axis), simulated for the cases of weak (N1), medium (N2-N4, typical for the Sun-like stars) and strong (N5) SW. According to these distributions, the collisional shock region between PW and SW in the case of high stellar mass loss rate (strong SW), approaches very close to the planet. This leads to shrinking of the absorbing area, populated by HeI($2^3$S), as shown in the left panel of Figure 2. According to the simulations, the projected size of HeI($2^3$S) absorbing area, or in other words, the distribution of the line-of-sight (LOS) absorption, as seen by remote Earth-based observer, appears to be rather small for GJ436b, as compared to the stellar disk size (right panel in Figure 3). This could be a reason why the transit absorption of GJ436b in line of metastable helium is rather tiny. At the same time, the projected size of the Lyα absorbing area (left panel in Figure 3) covers practically the whole stellar disk, resulting in significant Lyα absorption depth and well pronounced corresponding transit light curve.

To check the trends of the HeI($2^3$S) absorption change with the increasing/decreasing density of the SW, it is worth to look at the plots in the left panel of Figure 2, especially the distributions of HeI($2^3$S), atmospheric density and temperature, as well as at the 2D images of the absorbing area in Figure 4 (bottom panel), calculated for different SW densities. In particular, one can see in Figure 2 (left panel) the continuous broadening of the HeI($2^3$S) populated area below the ionopause with the decreasing density of the SW. This leads to small but monotonous increase of the total HeI($2^3$S) absorption, seen in Table 1 from N2 to N4 cases. The N1 case does not follow this trend because for this SW the shocked region ahead of the planet is beyond the stellar disk. However, in the N2 case it appears on the disk and a small increase in the density of HeI($2^3$S) (Fig. 2) gives a small increase in absorption. Note that while the difference in absorption between cases N1-N4 is only a factor of 1.3, the difference between N2 and N5 is a factor of 2. The plots of temperature and velocity distributions at the bottom of the left panel in Figure 2 enable better visualization of the absorbing area structure and size controlled by the ionopause borders under different SW conditions, as well as the relative position of the PW-SW interaction region (e.g. location of shock) and its influence on the absorbing area.

Thus, the medium SW conditions (model runs N2-N4) make no crucial influence on HeI($2^3$S) absorption because the PW-SW interaction region is located at relatively large distances from the planet. Only when the shocked region comes closer to the planet and He triplet is confined at x <= 3Rp (Figures 2 and 3), such Influence becomes pronounced. The comparison of the projected areas of absorption by HeI($2^3$S) for different strengths of the SW is provided in the bottom panel of Figure 4. In particular, it shows once again that under the conditions of weak and medium SW, the size of absorbing area remains approximately the same and weakly affected by the stellar plasma flow, while strong SW significantly decreases it.

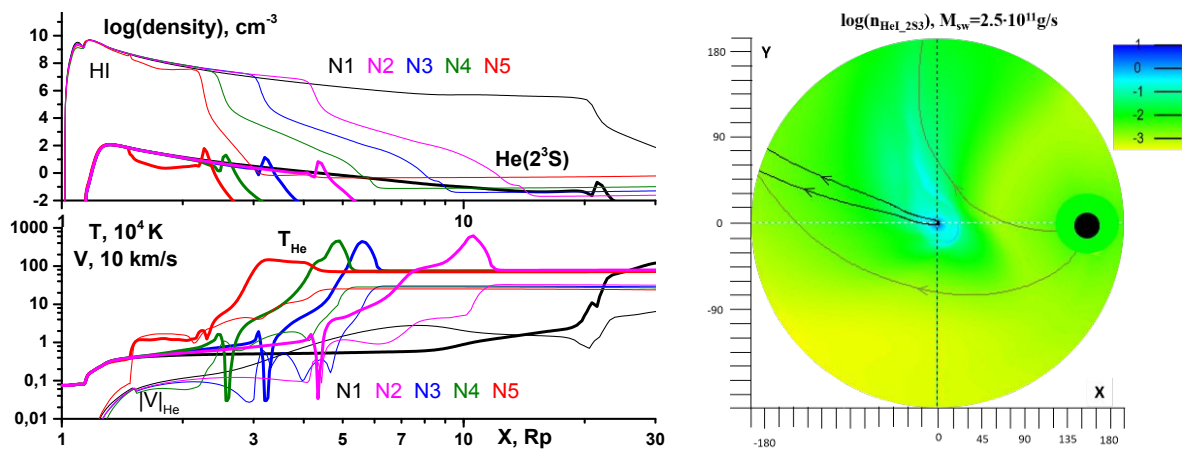

**Figure 2.** *Left panel*: Simulated in the model runs with weak (N1, black), medium (N2, N3, and N4, cyan, blue, and green) and strong (N5, red) SW the distributions along the planet-star line (X-axis) of density (upper frame, log scale) for atomic hydrogen (thin line) and metastable helium (thick lines), as well as temperature (bottom frame, thick line), and velocity (bottom frame, thin line). *Right panel*: Distribution of metastable helium in the equatorial plane (XY), simulated in the model run N1. Black and khaki streamlines show the flow of atomic hydrogen and protons, respectively. The planet is at the center of the coordinate system (0,0) and moves anti-clockwise around the star located at (158,0). The spatial scales, here and further on, are in units of planetary radii $R_p$.

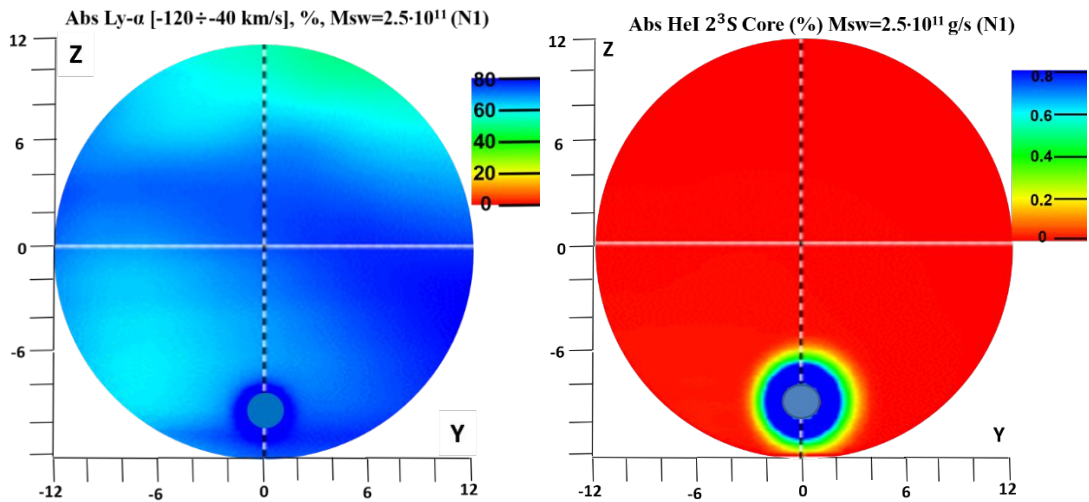

**Figure 3.** *Left panel*: Distribution of the LOS absorption by GJ436b, in Lyα, as seen by remote Earth-based observer at mid-transit, averaged over the blue wing ([-120; -40] km/s) of the line. *Right panel:* Distribution of the LOS

absorption by GJ436b, in HeI($2^3$S) triplet line across the stellar disk (the big circle). Note, that this picture uses the observer-based reference frame, in which the planet is shifted according to the inclination angle of the orbit to Z=-9 Rp relative the observer-star line (Z=0, X=0) (like in Khodachenko et al. 2019). All metric units expressed in planetary radius (Rp)

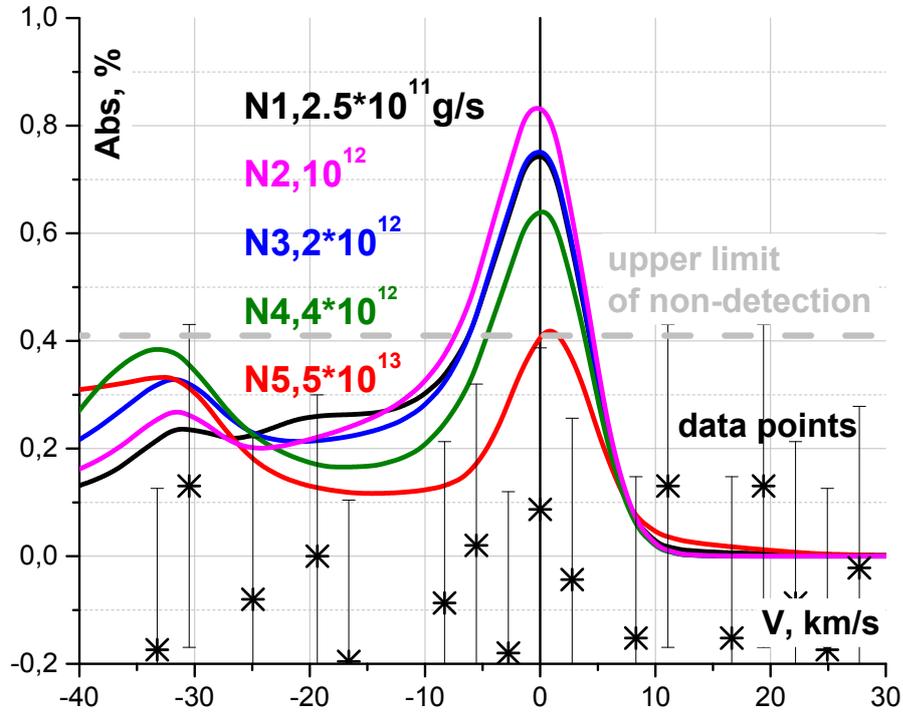

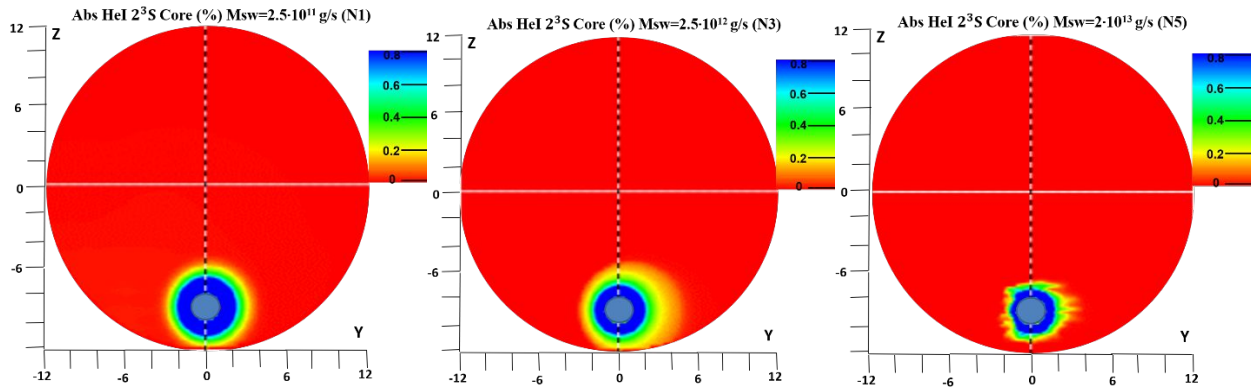

**Figure 4.** The effects of the SW on the HeI triplet population and absorption. *Upper panel:* HeI($2^3$S) absorption profiles calculated at mid-transit with different values of the stellar mass loss rate $\dot{M}_{sw}$, i.e. in the model runs with weak (N1, black), medium (N2, N3, and N4, cyan, blue, and green) and strong (N5, red) SW. The measured data *(Nortmann et al. 2018)* here and further in figures 5, 6 and 7 are shown with asterisks with error bars. *Bottom panel, left to right:* Distribution of the LOS absorption in HeI($2^3$S) triplet line at GJ436b as seen by remote Earth-based observer at mid-transit, simulated in the model runs with weak (N1), medium (N3) and strong (N5) SW, and overlaid above the stellar disk (the big circle).

Altogether, the performed study of the influence of SW variations on the transit absorption in HeI($2^3$S) triplet line at GJ436b leads to the conclusion that the moderate and weak SW have relatively weak effect on it. This result is similar to the outcome of analogous investigations of HD209458b in *Khodachenko et al. 2021* and HD189733b in *Rumenskikh et al. (2022)*. The absorbing metastable helium fraction is located relatively close to the planet and it is not affected by the specifics of interaction between SW and PW. At the same time, in the case of extreme SW the shrinkage and degradation of HeI($2^3$S) absorbing area takes place, resulting in overall decrease of the absorption. In general, the properties of this absorption cannot be directly used for deriving the SW parameters.

To check a possible role of some deviations of the stellar XUV flux relative the adopted in our basic case value $F_{XUV}$=0.86 erg·cm$^{-2}$·s$^{-1}$ at 1 a.u., the model runs with a twice lower as well as twice higher $F_{XUV}$ (N6 and N7, respectively), have been performed. As one can see in the right panel of Figure 5, the maximum absorption at the line center of HeI($2^3$S) increases with XUV flux, however not proportionally, due to the complex interplay of all reactions responsible for the population of metastable helium (see in Figure 1).

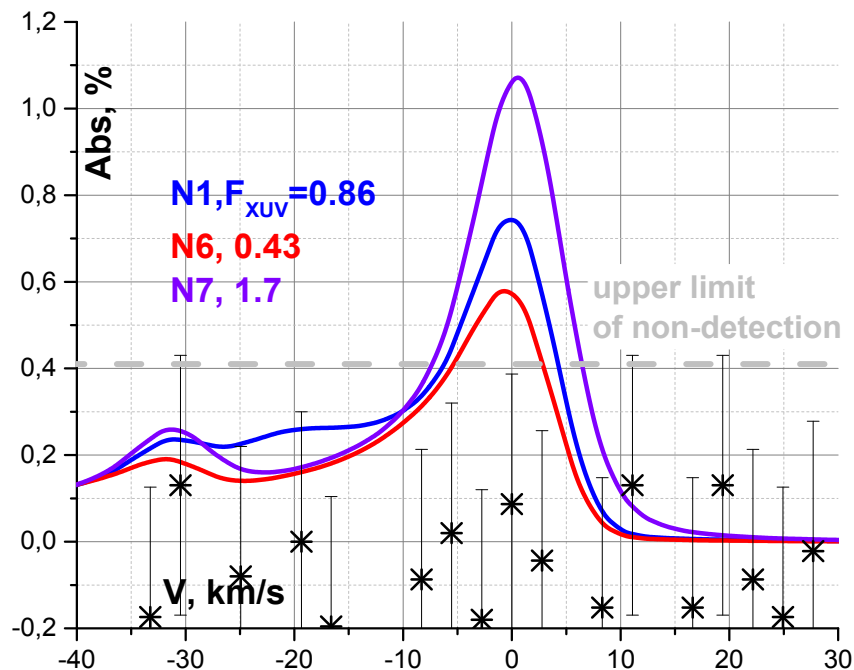

**Figure 5.** HeI($2^3$S) absorption profiles simulated with different XUV fluxes and fixed helium abundance He/H = 0.03 in the model runs N1, N6 and N7.

Altogether, in spite of the major role of recombination in the pumping of metastable HeI($2^3$S) level (Figure 1), the variations of ionizing XUV radiation flux have only moderate effect on the excess absorption at 10830Å line.

### *3.2 The effects of heavy element admixtures and helium abundance*

Since there are no evidences of secondary nature of the atmosphere on GJ436b, we assume its primordial type and take for our simulations the solar-like or sub-solar abundances of helium and those heavy trace elements, which have been detected in the transit absorption lines (i.e., SiIII, CII, *Dos Santos et al. 2019*). The metals could influence the processes in upper atmosphere due to reduction of molecular dissociation, hydrogen and helium ionization, because of catching part of the XUV radiation. This altogether should result in some decrease of the atmospheric escape. Moreover, possible presence of water in the atmosphere of planet may cause very rapid dissociation of $H_2$ molecules (*Garcia Muñoz et al. 2007, 2020*), resulting in the domination of atomic hydrogen in the base of the modelled atmosphere, instead of the usually assumed molecular content. This changes the scale-height and increases the density of upper atmosphere with the corresponding increase of mass loss.

To verify the above specified effects of the atmosphere composition on the material escape and overall mass loss, as well as absorption in metastable helium line, we performed a set of dedicated simulations with different abundances of helium and heavy trace elements, in particular, SiIII, CII, scaled with respect to the typical Solar system values (N11-N14), and considered also a special case with an atomic hydrogen dominated atmosphere (N10). The letter tests the hypothesis of possible dissociation of $H_2$ molecules by water.

The left panel of Figure 6 shows the mean absorption values in CII and SiIII lines, calculated from the corresponding simulated absorption profiles at mid-transit, as well as peak absorptions in HeI($2^3$S) line obtained for different abundances of the elements. Besides of that, we also provide here the comparison of helium peak absorptions, calculated in the cases of molecular (N1) and atomic hydrogen (N10) dominated atmospheres. As can be seen, significant absorption at the position of metastable HeI($2^3$S) takes place in the case of atomic hydrogen dominated atmosphere, which overcomes the upper limit of non-detection by almost an order and should be easily detected in the measurements. Absorption in CII line also overcomes the observational upper limit in this case. This is because the general increase of atmosphere density at distances of several $R_p$, results in 5 time increase of the outflow rate, as compared to that realized under the same conditions, but for the molecular hydrogen dominated atmosphere. We note that in *Oklopčić & Hirata (2018)* atomic base atmosphere was assumed with an assumed mass-loss rate of $2 \cdot 10^{10}$ g/s. It is sufficiently close to our modeled value of $3 \cdot 10^{10}$ g/s (run N10) but three times larger than for the molecular base atmosphere.

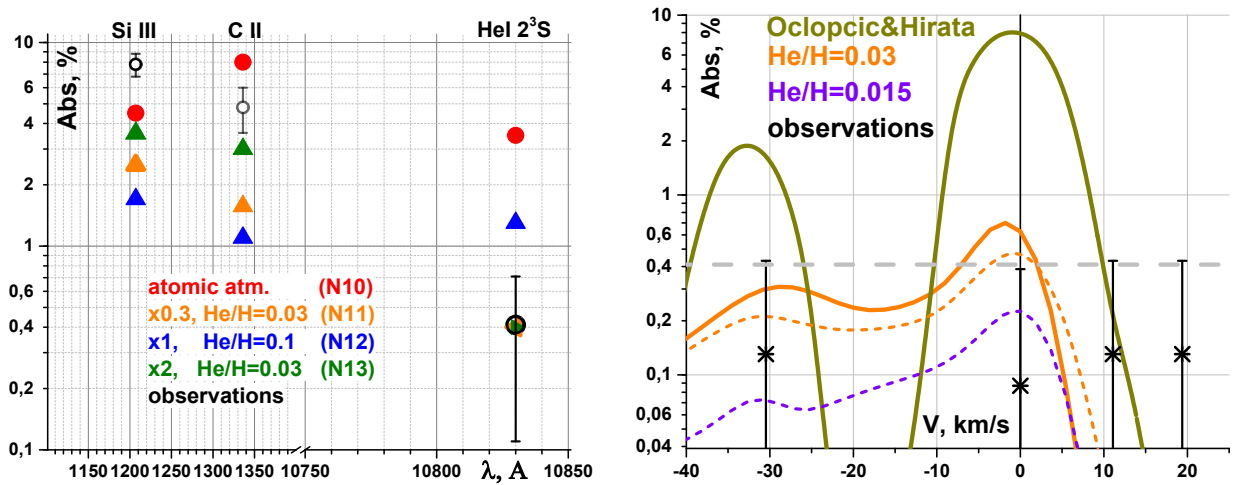

***Figure 6.*** *Left panel*: Simulated total absorptions in SiIII line at 1206.5 Å, CII triplet at 1335 Å and peak absorption values in HeI($2^3$S) line at 10830 Å, obtained in the model runs with different abundances of metals and helium for the molecular (N11-13) and atomic hydrogen (N10) dominated atmosphere. Open black circles with error bar denote the data imposed upper limit for transit absorptions at a 2σ confidence level by corresponding elements (metastable Helium according to *Nortmann et al. 2018*, CII and SiIII according to *Dos Santos et al. 2019*).

*Right panel*: Comparison of the observation data on HeI($2^3$S) absorption (asterisks) with the results of modeling in *Oklopčić & Hirata 2018* (khaki lines) and those obtained in our simulation run N1 (orange), with the space weather parameters adopted from *Khodachenko et al. 2019,* sub-solar helium content, and the realistic radiation pressure acting on metastable helium atoms. For more exact comparison with observations, we averaged the simulated absorption profile over the ±0.6 hours of transit (i.e., between the second and third contact points) shown with a dashed orange line. The dashed violet line shows the averaged absorption obtained with significantly sub-solar helium content He/H=0.015 (run N15).

According to the results of the model runs N11-N14, summarized in Table 1, the planetary mass loss rate in the case of molecular atmosphere is weakly dependent on the changes of the helium and heavy trace elements abundances, whereas the atmosphere absorption levels in the lines of the corresponding elements are nearly proportional to their content. Therefore, based on the simulated absorption levels shown in the left panel of Figure 6 in comparison to the measured values, we can constrain the helium and heavy elements in the atmosphere of GJ436b at the levels of ≤0.3 and ≤2 of the solar abundances, respectively. These provide the upper limits corresponding to upper limits of non-detection provided by observations. Moreover, our simulations with different abundances of heavy trace elements (N1, N11-N13) reveal that they make almost no effect on the HeI($2^3$S) absorption and the PW intensity. This is because the near-solar abundances of metals are rather low for the considered processes. As can be seen in the right panel of Figure 6, the largest helium abundance that fits observational non-detection limit is He/H=0.03. Note that to reproduce the conditions of data accumulation we plotted the simulated absorption profile averaged over the transit time, like in the observations reported in *Nortmann et al. 2018*. Figure 6 also shows absorption for the significantly sub-solar helium abundance of He/H=0.015 which falls below the non-detection level almost by a factor of two.

In regard of transit depths in CII and SiIII lines which we modeled in the present work, one can see that for solar-like abundances they are well below the upper-limits of non-detection derived by *Dos Santos et al. 2019*. Moreover, despite differences between our 3D model and that of 1D of *Loyd et al. 2017*, the results for CII line absorption are close enough.

### *3.3 The role of radiation pressure acting on HeI($2^3$S) atoms*

The near-IR radiation pressure acting on metastable helium atoms was shown to be unimportant for a number of exoplanets, except of the case of Wasp-107b, where it plays an important role in producing of the several km/s blue shift of absorption profile (*Khodachenko et al. 2021b*). To pinpoint the significance of near-IR radiation pressure on GJ436b, besides of the simulations with the measured flux value $F_{10830}$ = 3 erg s$^{-1}$ cm$^{-2}$ Å$^{-1}$ at 1 a.u. (N1), we performed the model runs N8 and N9 with the lower and higher near-IR fluxes, respectively, while keeping other model parameters the same. These simulations revealed that the near-IR radiation pressure force indeed affects the dynamics of HeI($2^3$S) atoms and the related absorption (see in Figure 7). In particular, in the case of typical flux, $F_{10830}$ = 3 erg s$^{-1}$ cm$^{-2}$ Å$^{-1}$ at 1 a.u., it exceeds the stellar gravity force by an order of magnitude and accelerates the HeI($2^3$S) atoms up to velocities of ~30 km/s (blue line in right panel of Figure 7). Note, that in the case of about 6 times higher NIR radiation flux (model run N9) and the correspondingly increased radiation pressure force $F_{rad}$, as compared to the real case, the velocity of HeI($2^3$S) atoms increases up to several hundreds of km/s. It can be estimated as $\Delta V = F_{rad} \cdot \tau / m_{He}$, where $\tau$ is the life-time of the metastable level. Since the value of $\tau$ can be rather high, and the radiation pressure force accelerates the HeI(23S) atoms in the wide range of their velocities (due to the continuity of the NIR spectrum around λ=10830Å), the achieved ΔV might be significant. Altogether, based on the modelling results, we can conclude that the account of the radiation pressure force produced by the real NIR flux of $F_{10830}$ = 3 erg s$^{-1}$ cm$^{-2}$ Å$^{-1}$ at 1 a.u. of the host star is certainly needed for the correct description of dynamics of HeI($2^3$S) atoms.

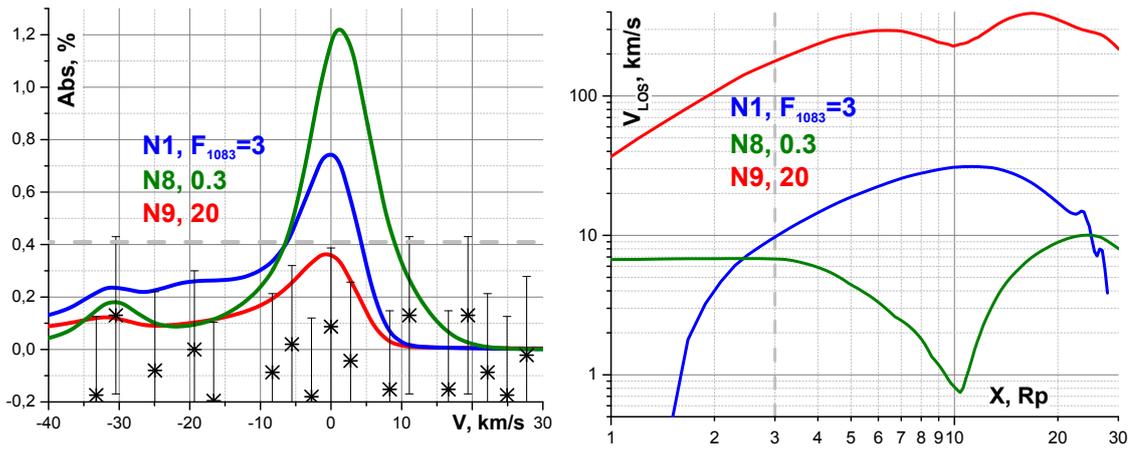

**Figure 7.** *Left panel*: HeI($2^3$S) absorption profiles, calculated in the model runs (N1, N8, and N9) with different $F_{10830}$ fluxes, i.e., different radiation pressure forces acting on HeI($2^3$S) atoms, versus observations. *Right panel*: Distribution of the helium atom velocity's projection along the LOS, parallel to the planet-star line (X-axis), calculated in the model runs N1, N8, and N9. The X axis is directed toward the star.

As can be seen in the left panel of Figure 7, the near-IR radiation pressure acting on the HeI($2^3$S) atoms affects not only the wings of absorption profiles, but the peak absorption values. When the radiation pressure is not taken into account, like in *Oklopčić & Hirata (2018)*, the simulated peak absorption in the line center is almost twice higher. In the case of higher than expected near-IR radiation flux the metastable helium atoms become more dispersed over a wider area along the LOS, resulting in the flattering of the absorption profile and further reduction of its peak value. Nevertheless, the stellar near-IR radiation flux is not a free parameter of the model and is well constrained by the stellar parameters and the corresponding measurements. At the same time, a small HeI($2^3$S) absorption peak at the level of ~0.4%, predicted by the simulations, might still remain undetected due to insufficient accuracy of the measurements.

**Conclusion**

The warm Neptune GJ436b is an object of special interest due to its extraordinary large transit absorption depth in Lyα line, produced by a huge hydrogen envelope, comparable in projection size with the stellar disk, as seen by an Earth-based observer. In the pioneering work by *Oklopčić & Hirata, (2018)* devoted to the study of HeI($2^3$S) absorption at GJ436b, using a simple semi-analytical 1D model, helium absorption profile of a quite detectable depth was calculated. However, this prediction turned to be rather different from the measurements. In this paper we presented an explanation why all the previous attempts of detection of the excess transit absorption at GJ436b in HeI($2^3$S) line, despite the intuitive expectations based on Lyα observations, were not successful.

The existed so far discrepancy between the modeling and observations highlights to our opinion the importance of an appropriate account in the numerical simulations of the whole complex of relevant physical processes, forces and geometrical aspects. Using our global 3D multi-fluid self-consistent numeric HD model which simulates in 3D escaping upper atmospheres of hot exoplanets taking into account, along with gravitational, centrifugal and hydrodynamic forces, the radiation pressure, SW, and basic chemistry of the atmospheric species, we demonstrated that the absorbing area, mostly populated by the metastable helium, is rather compact (<3$R_p$). Since the GJ436b is a rather small planet, the relatively compact projection of the HeI($2^3$S) absorbing area on the stellar disk naturally results in a tiny absorption. Thus, the main difference with the Lyα line absorption is that the fast hydrogen atoms producing high resonant absorption in Lyα are generated in a wide shocked layer quite far from the planet, while the metastable helium atoms are present in sufficient quantities only close to the planet.

A dedicated study was performed regarding the role of the near-IR radiation pressure force, acting on the metastable helium atoms, which is another reason of overestimating the HeI($2^3$S) absorption in simple 1D models. This force in the case of GJ436b overcomes the gravity force by an order of magnitude and causes certain dispersion and flattering of the HeI($2^3$S) absorption profile. Besides of using of the measured value of the NIR radiation flux ($F_{10830}$ = 3 erg s$^{-1}$ cm$^{-2}$ Å$^{-1}$ at 1 a.u.), we performed the model runs with the 10 times lower and about 6 times higher NIR fluxes (the model runs N8,9). This modelling has directly shown the influence of the radiation pressure on the motion of the metastable helium atoms and the corresponding shape/amplitude of the absorption profile. The performed simulations with different NIR fluxes demonstrate the importance of the account of the NIR radiation pressure for the case of GJ436b, as well as the importance of verifying it for other exoplanets.

We also simulated a scenario, when $H_2$ dissociation processes, intensified by possible presence of water in the atmosphere, result in the inflation of atmospheric envelope with the corresponding increase of the HeI($2^3$S) absorption, overcoming the observational limit. Therefore, the current non-detection of helium absorption might indicate about non-fulfillment of this scenario on GJ436b and therefore its mainly molecular hydrogen content of base atmosphere.

Our study of the contribution of heavy trace elements to the dynamics of escaping PW on GJ436b revealed that at the solar-like abundances it is rather small, and possible presence of metals is unlikely to influence the mass loss rate and actual non-detection of HeI($2^3$S) absorption. The absorption in SiIII, CII lines is about (1-2)% at standard abundances and well below of the available upper-limits of non-detection.

Similar to *Rumenskikh et al. 2022*, it was shown that the weak and medium SW, expected for GJ436b, insignificantly affect the compact absorbing area, populated by the metastable helium, and therefore their influence on the excess transit absorptions remains below the detection threshold.

Altogether, the performed extensive study of various possible reasons why the transit absorption in the metastable helium line was not observed for GJ436b, reveals that this is mainly due to the following three: 1) the small ratio between the compact absorbing area and stellar disk projections, as seen by the observer, 2) radiation pressure force acting on the HeI($2^3$S) atoms spreading them along the line of sight and around the planet, as well as 3) sufficiently small He/H ratio in atmosphere. The first reason also explains relatively small absorption in heavy trace elements. In regard of point 3, our simulations reveal that GJ436b atmosphere can contain low, but not exceedingly small amount of Helium, He/H≤0.3, to satisfy the currently available non-detection limit.

A general comment on the Helium abundance in upper atmospheres of exoplanets, as deduced from the observations at 10830 Å and undertaken numeric modelling, should be put forward. So far, in roughly half of the observed cases the non-detection of absorption at HeI($2^3$S) line takes place. Some of the non-detections are rather baffling (see the review in *Fossati et al. 2023*). At the same time, in the cases where HeI($2^3$S) absorption was detected, the modelling infers mostly sub-solar helium abundances He/H<0.03, e.g., in HD 209458 b (*Lampón et al. 2020; Khodachenko et al. 2021b*), HD 189733 b (*Lampón et al. 2021b; Rumenskikh et al. 2022*), GJ 3470 b

(*Shaikhislamov et al. 2021; Lampón et al. 2021b*), WASP 52b (*Yan et al. 2022*), GJ 1214b and HAT P 32 b (*Lampón et al. 2023*). Therefore, the constrain of the helium abundance at He/H≤0.3 revealed in the present paper, appears in line with this trend, showing that many exoplanets with the escaping atmospheres might indeed have helium abundances significantly smaller than that in the Solar system. However, there exists at least one case where a solar-like He/H value was derived, it is WASP 107 b (Khodachenko et al. 2021a). This means that the He/H ratio can be rather diverse in upper atmospheres of different exoplanets, and each particular case requires individual analysis and interpretation, taking into account the specific features of the considered stellar-planetary systems. Anyway, as shown in the modelling of GJ436 b reported here, a non-detection of helium triplet absorption in particular exoplanet should not be ad hoc treated as an indication of absence of helium in its upper atmosphere. Complex dynamics of the escaping atmospheric material, interacting with the surrounding stellar wind, and varying size of the HeI($2^3$S) absorbing region are unique for each planet and have to be considered accordingly. The reported here simulations of GJ 436 b and the available measurements, enable inferring only an upper limit of the helium abundance at the planet, but not its actual value.

*Acknowledgements.* The work was supported by RSF grant 23-12-00134. MLK acknowledges the projects I2939-N27 and S11606-N16 of the Austrian Science Fund (FWF).

**Data Availability**

The data underlying this article will be shared on reasonable request to the corresponding author.